\documentclass[10pt]{iopart}
\usepackage[citestyle=numeric-comp,bibstyle=numeric]{biblatex}
\usepackage[utf8]{inputenc}
\expandafter\let\csname equation*\endcsname\relax
\expandafter\let\csname endequation*\endcsname\relax
\usepackage{mathtools}
\usepackage{bbm}
\usepackage{iopams}  
\usepackage{bm}
\usepackage{xparse}
\usepackage{relsize}
\newcommand{\re}{\mathbb{R}}

\renewcommand{\l}{\left}
\renewcommand{\r}{\right}

\newcommand{\bb}{\mathbf{b}}
\newcommand{\cb}{\mathbf{c}}

\newcommand{\nb}{\mathbf{n}}

\newcommand{\ssb}{\mathbf{s}}

\newcommand{\xb}{\mathbf{x}}

\newcommand{\Gammab}{\mathbf{\Gamma}}
\newcommand{\Xib}{\mathbf{\Xi}}
\newcommand{\zetab}{\boldsymbol{\zeta}}
\newcommand{\xib}{\boldsymbol{\xi}}
\newcommand{\Bb}{\mathbf{B}}

\newcommand{\Eb}{\mathbf{E}}

\newcommand{\Ob}{\mathbf{O}}

\newcommand{\Rb}{\mathbf{R}}

\let\temp\phi
\let\phi\varphi
\let\varphi\temp

\newcommand{\alphab}{\boldsymbol{\alpha}}

\newcommand{\vertiii}[1]{{\left\vert\kern-0.25ex\left\vert\kern-0.25ex\left\vert #1 
    \right\vert\kern-0.25ex\right\vert\kern-0.25ex\right\vert}}

\def\Proj_#1{\ensuremath{\operatorname{Proj}_{#1}\!}}

\newcommand{\dr}[1]{\ensuremath{\operatorname{d}\!{#1}}}

\def\drs_#1{\ensuremath{\operatorname{d}_{#1}\!}}
\def\Drs_#1{\ensuremath{\operatorname{D}_{#1}\!}}
\def\Drs^#1{\ensuremath{\operatorname{D}^{#1}\!}}

\def\Ers_#1{\mathrm{E}_{#1}}

\def\xunderbrace#1_#2{{\underbrace{#1}_{\mathclap{#2}}}}
\def\xoverbrace#1^#2{{\overbrace{#1}^{\mathclap{#2}}}}
\def\xunderarrow#1_#2{{\underset{\overset{\uparrow}{\mathclap{#2}}}{#1}}}
\def\xoverarrow#1^#2{{\overset{\underset{\downarrow}{\mathclap{#2}}}{#1}}}

\DeclareDocumentCommand{\boxedeq}{m o}{%
    \IfNoValueTF{#2}{%
        \rlap{\boxed{#1}}%
        \phantom{\hskip\fboxrule\hskip\fboxsep #1}%
        }{%
        \rlap{\boxed{#1#2}}%
        \phantom{\hskip\fboxrule\hskip\fboxsep #1}&\phantom{#2}%
    }%
}

\renewcommand{\P}[1]{
    \mathbb{P}\l(#1\r)
}

\usepackage{standalone}
\usepackage{xcolor}
\definecolor{blue}{rgb}{0.2980392156862745, 0.4470588235294118, 0.6901960784313725}
\definecolor{green}{rgb}{0.3333333333333333, 0.6588235294117647, 0.40784313725490196}
\definecolor{red}{rgb}{0.7686274509803922, 0.3058823529411765, 0.3215686274509804}
\definecolor{purple}{rgb}{0.5058823529411764, 0.4470588235294118, 0.6980392156862745}
\definecolor{yellow}{rgb}{0.8, 0.7254901960784313, 0.4549019607843137}
\definecolor{lightblue}{rgb}{0.39215686274509803, 0.7098039215686275, 0.803921568627451}
\definecolor{orange}{rgb}{0.8666666666666667, 0.5176470588235295, 0.3215686274509804}
\definecolor{brown}{rgb}{0.5764705882352941, 0.47058823529411764, 0.3764705882352941} \definecolor{pink}{rgb}{0.8549019607843137, 0.5450980392156862, 0.7647058823529411}
\definecolor{gray}{rgb}{0.5490196078431373, 0.5490196078431373, 0.5490196078431373}
\usepackage{tikz}
\usepackage{expl3}
\usepackage{siunitx}
\usetikzlibrary{calc,arrows}
\usepackage{xspace}
\usepackage{amsmath}           
\usepackage{amssymb}
\usepackage{float}
\usepackage{booktabs}
\usepackage{pgfplots}
\usepackage{pgfplotstable}
\usetikzlibrary{pgfplots.groupplots}
\pgfplotsset{compat=1.13}
\newcommand{\lmax}{L_{\max}} \newcommand{\dmin}{d_{\min}}
\newcommand{\msc}{\mathrm{msc}}
\usepackage{todonotes}
\usepackage{amsthm}
\newtheorem*{remark}{Remark}
\usepackage{enumitem}

\usepackage{siunitx}

\newcommand{\rev}[1]{\textcolor{black}{#1}}

\begin{document}
\providecommand{\datadir}{./plots/}%
\title[Increasing tolerances in coil design]{Stochastic and a posteriori optimization to mitigate coil manufacturing errors in stellarator design}
\author{Florian Wechsung$^1$, Andrew Giuliani$^1$, Matt Landreman$^2$, Antoine Cerfon$^1$, Georg Stadler$^1$}
\address{$^1$ Courant Institute of Mathematical Sciences, New York University, New York, USA}
\address{$^2$ Institute for Research in Electronics \& Applied Physics, University of Maryland-College Park, Maryland, USA}
\ead{wechsung@nyu.edu, giuliani@cims.nyu.edu, mattland@umd.edu, cerfon@cims.nyu.edu, stadler@cims.nyu.edu}
\date{January 2022}

\begin{abstract}
    It was recently shown in [Wechsung et.~al., Proc.~Natl.~Acad.~Sci.~USA,
    2022] that there exist electromagnetic coils that generate
    magnetic fields which are excellent approximations to quasi-symmetric
    fields and have very good particle confinement properties. Using a Gaussian
    process based model for coil perturbations, we investigate the impact of
    manufacturing errors on the performance of these coils. We show that even
    fairly small errors result in noticeable performance degradation. \rev{While 
    stochastic optimization yields minor improvements, it is not able to mitigate these errors significantly.
    As an alternative to stochastic optimization,} we then formulate a new optimization problem for computing optimal
    adjustments of the coil positions and currents without changing the shapes
    of the coil. These a-posteriori adjustments are able to reduce the impact
    of coil errors by an order of magnitude, providing a new perspective for
    dealing with manufacturing tolerances in stellarator design.
\end{abstract}

\section{Introduction}
\begin{figure}
    \begin{center}
            {\includegraphics[height=5.3cm]{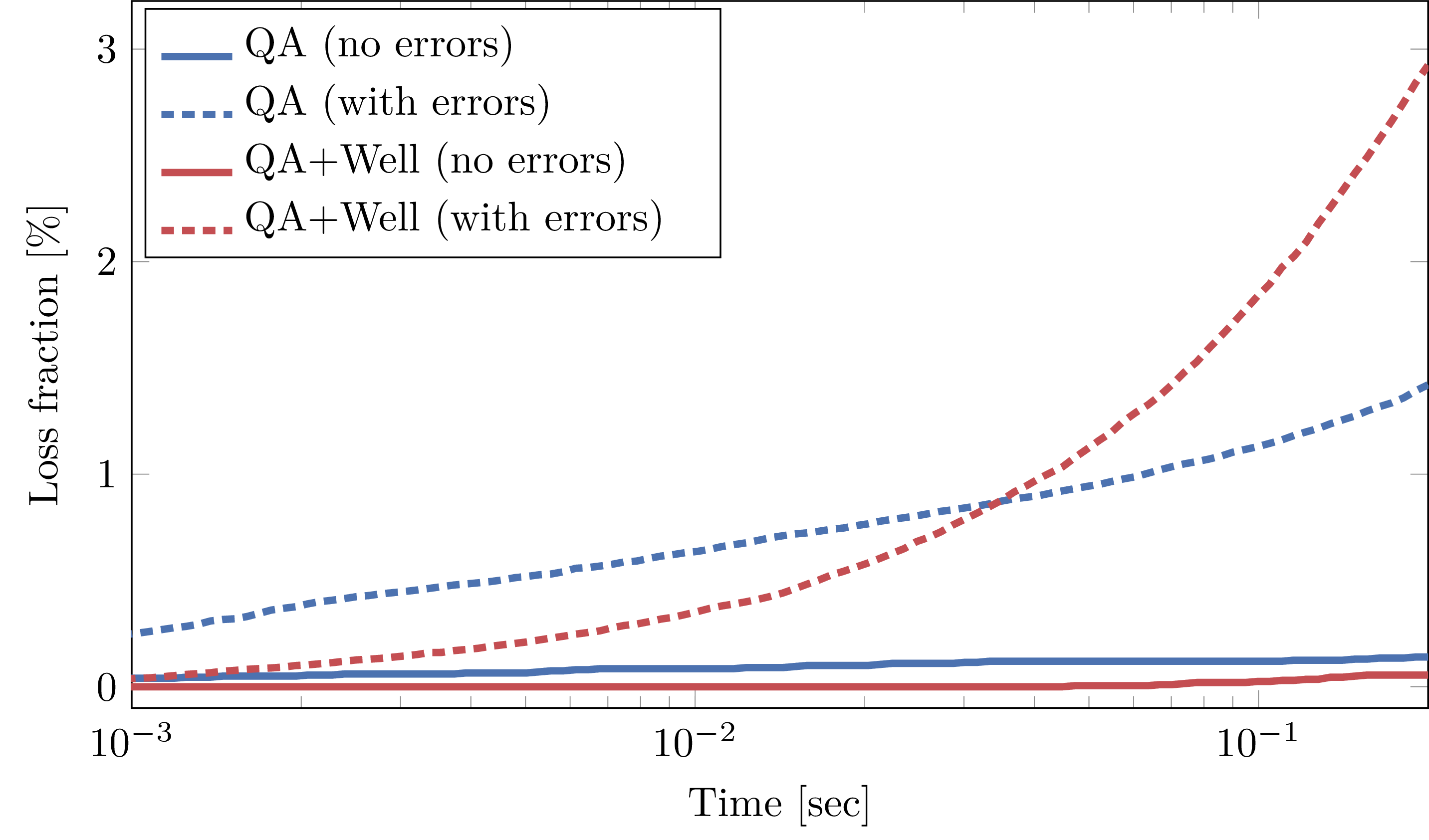}}
    \end{center}
    \caption{
        Loss fraction of alpha particles spawned at half-radius over time. Coil errors are approximately 1000 times smaller than the diameter of the device.
        We observe that such coil perturbations result in significant degradation of particle confinement.
}\label{fig:intro-conf}
\end{figure}

One of the most expensive and technically challenging aspects of the
construction of stellarators \cite{Strykowsky09, Neilson10,Klinger2013} is the
design, manufacturing and assembly of the primary coil system that generates the
magnetic field confining the plasma.
As computing resources and simulation capabilities have increased over the last
few decades, large-scale numerical optimization has proven to be the method of
choice for the design of new stellarator experiments~\cite{hegna_improving_2022}.
Numerical simulations of guiding-center trajectories have shown that the state-of-the-art stellarator designs obtained from these optimization efforts are now able to confine even highly energetic particles for long timescale with minimal losses --- a key requirement for practical use as fusion reactors~\cite{bader_advancing_2020,landreman_magnetic_2021,wechsung2022precise,giuliani2022}.
However, good performance of a stellarator can only be guaranteed if the coils are manufactured and placed with high accuracy~\cite{andreeva_analysis_2004,hammond_experimental_2016,shimizu_consideration_2019},
a requirement that has turned out to be a significant cost driver and risk factor for any plans to build a stellarator experiment~\cite{Strykowsky09,Neilson10}.
To illustrate the fragility of confinement to even rather small coil errors, we consider coils~\cite{wechsung2022precise} that approximate the two quasi-axisymmetric \rev{(QA)} configurations found in~\cite{landreman_magnetic_2021}.
We compute particle losses for alpha particles for the case when the coils are built perfectly, and when they have errors of a few millimeters, which may be compared to coil lengths of the order of a few meters (we give a precise description of the error model in Section~\ref{sec:formulation}), cf.~Figure~\ref{fig:intro-conf}. We observe that particle losses have a significant relative increase  in the presence of errors: from $0.14\%$ to $1.56\%$ for the design without a magnetic well, and from $0.06\%$ to $3.05\%$ for the design with a magnetic well.
While these numbers remain small in absolute terms compared to previous stellarators (\cite{landreman_magnetic_2021}, Figure 6a), the increased losses are clearly undesirable, and may be larger than acceptable level for a steady-state fusion reactor \cite{iter1999technical,Windsor1999}.

In an effort to lessen the impact of errors, stochastic optimization approaches
have recently drawn significant attention in the stellarator community~\cite{lobsien_stellarator_2018, lobsien_physics_2020, lobsien_improved_2020, wechsung2021, glas_2021}.
The core idea underlying the various approaches is the following: by including coil
errors explicitly as part of the optimization problem, one hopes to find
minimizers that perform well even when the coils are not built or placed
perfectly. Using this approach, Lobsien et al.~\cite{lobsien_stellarator_2018} were able to
find coils for the W7-X experiment that on average improve their coil design
objective by 20\% compared to those obtained from standard deterministic optimization. Accordingly, a significant fraction of this article is focused on the merits and limitations of stochastic optimization for coil design.  

While the methods we propose are generally applicable, 
in this work we focus on the vacuum magnetic fields found by Landreman \& Paul~\cite{landreman_magnetic_2021}.
These fields satisfy quasi-symmetry to a formerly unprecedented accuracy and have excellent confinement properties.
Furthermore, it was recently shown in~\cite{wechsung2022precise} that it is possible to design electromagnetic coils that reproduce these fields to high precision.
However, as these calculations did not take coil errors into account, the goal
of this present work is to understand, and then mitigate, the impact of such errors
on the performance of the stellarator. \rev{This work contains two main contributions. First}, we show that, perhaps surprisingly, stochastic
optimization only yields marginal improvements, and we provide an explanation for these results in the context of our optimization problem.  \rev{Second,} we show that, on the other hand, an a-posteriori correction
approach based on minor adjustments of the coils \emph{after} they have been
built is able to mitigate manufacturing errors to a \rev{significant} degree.
The idea behind our approach is similar to the coil correction process
performed for the W7-X experiment during
construction~\cite{andreeva_influence_2009,bosch_technical_2013}, but differs
in the choice of objective function and also considers adjustments to the
currents. 

The manuscript is structured as follows:
using the model for coil errors introduced in~\cite{wechsung2021}, we first perform a systematic study of the performance of the confining magnetic field as the magnitude of the error is increased.
We find that the precise quasi-symmetry property is quickly lost even for small coil errors.
In order to improve robustness, we then apply stochastic optimization to the problem.
While this results in slightly smaller average field error, the effect is small and does not result in improved particle confinement. 
Motivated by this finding, we propose a novel strategy to deal with coil manufacturing errors.
Assuming that the coils (and hence the errors that were made during production) can be measured accurately before placing them,
we solve a separate optimization problem to find adjustments to the placement of the coils that mitigate manufacturing errors as much as possible.
Assuming perfect measurements and placement, we show that these position adjustments are able to reduce the impact of manufacturing errors by several orders of magnitude.
\rev{This result motivates a larger emphasis on the accurate measurement and placement of coils, at a level comparable to the attention given to their manufacturing.}

\section{Stochastic optimization for precise quasi-symmetry}\label{sec:formulation}
\subsection{Formulation of the coil optimization problem and model for coil errors}
Following the approach of the FOCUS coil design tool~\cite{zhu2017}, we solve an optimization problem to find coil shapes and currents so that the magnitude of the normal component of the magnetic field induced by the coils is minimized on the boundary $S$ of a given ideal MHD equilibrium.
We use the same setup as in~\cite{wechsung2022precise}: for each of the two quasi axisymmetric fields of~\cite{landreman_magnetic_2021}, 16 coils are optimized in order to reproduce the target field.
Since the field satisfies two-fold rotational symmetry as well as stellarator symmetry, we only need to describe the shape and current of four independent modular coils to determine the whole system.
We make the common single filament assumption and represent each coil as a smooth, periodic function $\Gammab^{(i)}:[0,2\pi]\to\re^3$, represented by a truncated Fourier series \cite[eqn. (2)]{zhu2017}.

We consider an objective that is given by the \rev{relative} quadratic flux on the target surface,
\begin{equation}\label{eqn:fluxobjective}
    f(\Bb) = \int_S \bigg(\frac{\Bb\cdot \nb}{\|\Bb\|}\bigg)^2 \dr s,
\end{equation}
and denote the magnetic field induced by a set of coils by $\Bb(\cb) = \Bb(\{\Gammab^{(i)}(\cb)\}_{i=1}^{16}, \{I^{(i)}(\cb)\}_{i=1}^{16})$, i.e., the parameter vector $\cb$ contains the Fourier coefficients describing the four independent coils and their currents $I^{(i)}(\cb)$, and the remaining 12 coils and currents are determined through symmetries.
We recall that if $f(\Bb)=0$, then the field induced by the coils matches the target field exactly in the volume bounded by the surface.

Simply minimizing $\cb \mapsto f(\Bb(\cb))$ is an ill-posed problem in general, so in order to obtain coils that can be constructed at reasonable cost, the flux objective is usually combined with penalties or constraints on the geometry of the coils.
In this work, we follow the formulation of~\cite{wechsung2022precise} and consider inequality constraints on the gap between coils, their mean squared and maximum curvature, and their length.
Shorter coils are usually less complex geometrically, which is why the coil length is a commonly used regularizer in coil optimization.
In order to study the sensitivity of optima to this regularizer, we vary the coil length constraint, which bounds the sum of the lengths of the four independent coils, from $\lmax=\SI{18}{\meter}$ to $\lmax=\SI{24}{\meter}$ to obtain several coil sets of varying complexity and performance.
These lengths are relative to a device configuration scaled to have an average major radius of $\SI{1}{\meter}$.
We refer to a specific coilset using the notation `Target[$\lmax$]', e.g., QA[24] refers to the coilset that targets the QA configuration of~\cite{landreman_magnetic_2021} and for which the four independent modular coils have a combined length of maximally \SI{24}{\meter}.

As demonstrated in~\cite{wechsung2022precise}, solving this optimization problem yields coils that achieve very precise quasi-symmetry and low particle losses.
However, for the design of a practical experiment or reactor it is crucial to understand the impact that coil errors have on the magnetic field and on its ability to confine particles.
In order to answer this question, we require a model for coil errors that arise during manufacturing and placement.
Following the approach of~\cite[Section 2]{wechsung2021}, we denote the coil errors by $\Xib^{(i)}$ and consider the perturbed coils with parameterization $\Gammab^{(i)} + \Xib^{(i)}$.
We model coil errors as the sum of a systematic error, which satisfies two-field-period symmetry and stellarator symmetry, and statistical error, which is independent for each coil, that is $\Xib^{(i)} = \Xib^{(i)}_{\text{sys}} + \Xib^{(i)}_{\text{stat}}$.
The two different sources of error are modelled using a Gaussian process with the same pointwise standard deviation $\sigma$.
To give intuition for the resulting magnitude of the error, we consider the normal component of the perturbation.
Denoting by $\nb$ and $\bb$ a normal and binormal vector along the curve, we note that at each point, $P(\theta) = \|(\Xib^{(i)}(\theta) \cdot \nb(\theta))\nb(\theta) + (\Xib^{(i)}(\theta) \cdot \bb(\theta))\bb(\theta)\|$ is Rayleigh($\sqrt{2}\sigma$) distributed, and hence the expected pointwise normal error magnitude is $\Eb(P(\theta)) = \sqrt{\pi}\sigma$.
The distribution of the maximum error along the coil is more complicated, but in our experiments it is usually a small multiple of the expected normal error magnitude.

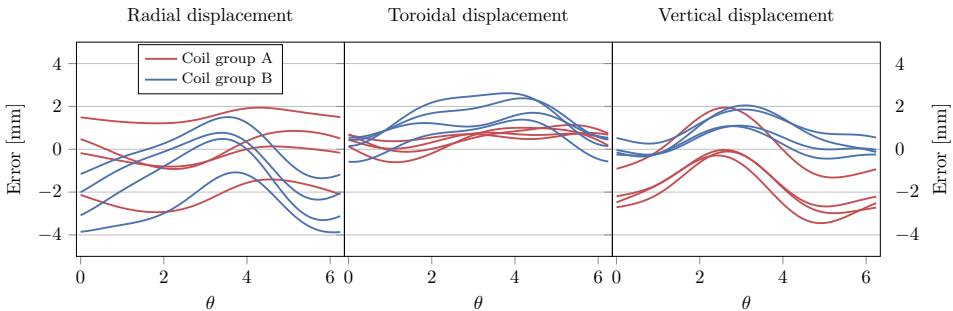
\begin{figure}[H]
    \begin{center}
        \providecommand{\datadir}{.}%
\pgfplotstableread[col sep = semicolon, header=true]{\datadir/perturbation.txt}\perturbationdata

\pgfplotsset{style1/.style={solid, line width=1.0pt}}
\pgfplotsset{style2/.style={dashed, line width=1.0pt}}
\pgfplotsset{style3/.style={dotted, line width=1.0pt}}

\begin{tikzpicture}
    \pgfplotsset{scaled y ticks=false}
    \pgfplotsset{scaled x ticks=false}
    \begin{groupplot}[
        group style={
            group name=my plots,
            group size=3 by 1,
            vertical sep=1.5cm,
            horizontal sep=0cm
        },
        width=5cm,
        height=4cm,
        tickpos=left,
        xmin=-0.1, xmax=6.3412,
        ylabel near ticks,
        legend cell align=left,
        legend style={at={(0.50, 0.99)}, anchor=north},
        legend columns=1,
        legend style={font=\footnotesize},
        scale only axis,
        ymin=-5,
        ymax=+5,
        scale only axis,
        ylabel style={overlay}, yticklabel style={overlay}, 
        trim axis left,
        xticklabel style={
            /pgf/number format/fixed,
            /pgf/number format/precision=4
        },
        scaled x ticks=false,
        ]
      \nextgroupplot[ymajorgrids=true,ytick align=outside, xtick align=outside, xlabel={$\theta$}, ylabel={Error [\SI{}{\mm}]}, title={Radial displacement}]
      \addplot[style1, red] table[x=phi,y=err_r_0_0] {\perturbationdata};
      \addlegendentry{Coil group A}
      \addplot[style1, red, forget plot] table[x=phi,y=err_r_1_0] {\perturbationdata};
      \addplot[style1, red, forget plot] table[x=phi,y=err_r_2_0] {\perturbationdata};
      \addplot[style1, red, forget plot] table[x=phi,y=err_r_3_0] {\perturbationdata};
      \addplot[style1, blue] table[x=phi,y=err_r_0_1] {\perturbationdata};
      \addlegendentry{Coil group B}
      \addplot[style1, blue] table[x=phi,y=err_r_1_1] {\perturbationdata};
      \addplot[style1, blue] table[x=phi,y=err_r_2_1] {\perturbationdata};
      \addplot[style1, blue] table[x=phi,y=err_r_3_1] {\perturbationdata};
      \nextgroupplot[ymajorgrids=true,ytick align=outside, xtick align=outside, xlabel={$\theta$}, ylabel={}, title={Toroidal displacement}, yticklabels={,,}]
      \addplot[style1, red] table[x=phi,y=err_phi_0_0] {\perturbationdata};
      \addplot[style1, red] table[x=phi,y=err_phi_1_0] {\perturbationdata};
      \addplot[style1, red] table[x=phi,y=err_phi_2_0] {\perturbationdata};
      \addplot[style1, red] table[x=phi,y=err_phi_3_0] {\perturbationdata};
      \addplot[style1, blue] table[x=phi,y=err_phi_0_1] {\perturbationdata};
      \addplot[style1, blue] table[x=phi,y=err_phi_1_1] {\perturbationdata};
      \addplot[style1, blue] table[x=phi,y=err_phi_2_1] {\perturbationdata};
      \addplot[style1, blue] table[x=phi,y=err_phi_3_1] {\perturbationdata};
      \nextgroupplot[ymajorgrids=true,ytick align=outside, xtick align=outside, xlabel={$\theta$}, ylabel={Error [\SI{}{\mm}]}, title={Vertical displacement}, yticklabel pos=right]
      \addplot[style1, red] table[x=phi,y=err_z_0_0] {\perturbationdata};
      \addplot[style1, red] table[x=phi,y=err_z_1_0] {\perturbationdata};
      \addplot[style1, red] table[x=phi,y=err_z_2_0] {\perturbationdata};
      \addplot[style1, red] table[x=phi,y=err_z_3_0] {\perturbationdata};
      \addplot[style1, blue] table[x=phi,y=err_z_0_1] {\perturbationdata};
      \addplot[style1, blue] table[x=phi,y=err_z_1_1] {\perturbationdata};
      \addplot[style1, blue] table[x=phi,y=err_z_2_1] {\perturbationdata};
      \addplot[style1, blue] table[x=phi,y=err_z_3_1] {\perturbationdata};
    \end{groupplot}
    \node at (16.5cm, 0cm)(a) {};
    \node at (-1.5cm, -1cm)(a) {};
\end{tikzpicture}
    \end{center}
    \caption{Samples drawn from the error model for two \rev{coil} groups \rev{(A and B)}. Shown in the same colour are samples from our error \rev{model} for coils related through stellarator and two-field-period symmetry. We see that part of the error is the same across the four instances of the modular coils, \rev{i.e., within group A and group B}. \rev{The reader may compare these figures with the measurements for W7-X in \cite[Figure~3]{Andreeva_2014}.}}\label{fig:perturbation}
\end{figure}

The length scales for the stochastic processes are chosen to be $\pi/2$ for the systematic error, and $\pi$ for the statistical error.
An example for a coil perturbation drawn from this model can be seen in Figure~\ref{fig:perturbation}; \rev{in magnitude and structure} this error is similar to that measured for the W7-X experiment \cite[Figure~3]{Andreeva_2014}.

We collect the information about the Gaussian processes underlying the deformations in $\zetab$ and denote the magnetic field generated by a set of perturbed coils by
\begin{equation}
    \Bb(\cb, \zetab) = \Bb\big(\{\Gammab^{(i)}(\cb) + \Xib^{(i)}(\zetab)\}_{i=1}^{16}, \{I_i(\cb)\}_{i=1}^{16}\big).
\end{equation}
The stochastic optimization problem that we solve is then given by
\begin{subequations}\label{eqn:optimproblem}
    \begin{align}
        \underset{\cb}{\mathrm{minimize}}\quad &   \Eb_{\zetab}\Big[f\big(\Bb(\cb, \zetab)\big)\Big],\\
        \text{subject to}     \quad&   \mathrm{Max}(\kappa_i)  \le \kappa_{\max} \text{ for } i=1,\ldots,4,\label{seqn:maxkappa}\\
                                   &   \mathrm{Mean}(\kappa_i^2)  \le \kappa_{\msc} \text{ for } i=1,\ldots,4,\label{seqn:meankappa}\\
                                   &   \sum_{i=1}^{4} L_i   \le \lmax, \label{seqn:len}\\
                                   &   \min_{\theta,\theta'} \| \Gammab_i(\theta)-\Gammab_j(\theta')\|   \ge \dmin\text{ for all } 1\le i < j \le 16,\label{seqn:dist}\\
                                   &    \|\Gammab_i'\| = \frac{L_i}{2\pi} \text{ for } i=1,\ldots,4.\label{seqn:arclen}
    \end{align}
\end{subequations}
Here~\eqref{seqn:maxkappa} limits the maximum curvature along each coil, \eqref{seqn:meankappa} limits the mean-squared curvature along each coil, \eqref{seqn:len} limits the combined length of the independent modular coils, and~\eqref{seqn:dist} enforces that the coils are separated from each other.
Finally, the constraint in~\eqref{seqn:arclen} ensures a constant incremental arclength in the parametrization of the coils. We include this constraint for two reasons:
\begin{enumerate}
    \item To ensure differentiability of line integrals: the mean squared
        curvature for example is given by $\frac{1}{L_i} \int_{[0,2\pi)}
        \kappa_i(\theta) \|\Gammab'_i(\theta)\| \dr \theta$. The incremental
        arclength term in that integral is not differentiable with respect
        to coil coefficients if $\|\Gammab'_i(\theta)\|=0$ anywhere.
    \item To ensure constant length scale of the Gaussian process error model: if the incremental
        arclength is allowed to vary along the coil, then this might distort
        the characteristic length scale of the Gaussian process. In
        experiments without this constraint, we observe that the optimization
        algorithm exploits this freedom. This results in a lower objective
        value, deceptively implying that stochastic optimization has larger
        benefit than it actually does. In reality, some of the improvement in
        the objective is due to varying the arclength around the curve.  This
        causes quadrature points along the coil to bunch together or to separate
        from one another, thereby nonphysically smoothing the error in some
        regions, or making it highly oscillatory in others. This is not
        desirable as the representation of the curve should not impact the
        nature of the errors.
\end{enumerate}

To approximate the expectation over the coil errors, we draw samples $\zetab_k$ of the error distribution, evaluate the flux objective for each perturbed set of coils and compute the Monte Carlo sample average, that is
\begin{equation}
    \Eb_{\zetab}[f(\Bb(\cb, \zetab))] \approx \frac{1}{N} \sum_{k=1}^N f(\Bb(\cb, \zetab_k)). \label{eqn:sampleavg}
\end{equation}
The deterministic optimization problem can be recovered by simply setting $\zetab=0$.

We solve this optimization problem using the SIMSOPT software~\cite{simsopt}.
SIMSOPT is an open-source library for stellarator optimization that is mostly written in Python, with performance-critical parts (such as the Biot-Savart law) implemented in C++.
The constraints are implemented as penalty functions, and we increase each penalty term until the constraint violation is less than $0.1\%$, as also done in~\cite{wechsung2022precise}.
Derivatives of the flux objective and the regularization terms are computed by SIMSOPT using a mix of manual implementation and automatic differentiation based on JAX~\cite{jax2018github}.
SIMSOPT uses the L-BFGS~\cite{nocedal1980} quasi-Newton scheme to solve the optimization problem efficiently. L-BFGS uses a history of gradients evaluated at different parameters to approximate second derivatives, which we found to be important for convergence as it helps to overcome the ill-posedness of the optimization problem.
We run the L-BFGS iteration for 14\,000 iterations and then perform an additional five steps of Newton's method, using a finite difference approximation for the Hessian, in order to ensure convergence.
For the shorter coils the problem is better conditioned and L-BFGS on its own is able to reduce the norm of the gradient of the objective by 10 orders of magnitude or more.
For the longer coils, L-BFGS is only able to reduce the gradient by about seven orders of magnitude and we gain an extra order of magnitude reduction using the Newton iterations.
We note that poor conditioning of coil design problems as the coil length increases was also observed in~\cite[Fig.~4]{giuliani2022}.

To mitigate issues of local minima, we ran the deterministic optimization for 8 different initial guesses and picked the best performing configuration, as measured by the flux objective.
The different initial guesses are obtained by performing random perturbations of regularly spaced circular coils.
For the stochastic optimization, we run 16 optimizations: 8 that use the minimizers of the deterministic problem as initial guesses, and 8 that use perturbations of circular coils as initial guess.
We use 4096 samples to approximate the mean in~\eqref{eqn:sampleavg}.
Each stochastic optimization problem is run on 128 cores across eight nodes consisting of two Intel Xeon Platinum 8268 CPUs and takes approximately eight hours to solve.

\subsection{Numerical results}\label{sec:stoch-num}
First, we evaluate the impact of stochastic optimization directly on the flux objective. Then, we additionally study how designs obtained with the deterministic and the stochastic approaches perform with respect to physical quantities of interest such as alpha particle confinement. 

We solve the stochastic problems for perturbations with $\sigma=\SI{1}{mm}$, as well as the deterministic problem.
For the obtained coil designs, we then draw 128 new perturbations of magnitude ranging from $\sigma=\SI{0}{\mm}$ to $\sigma=\SI{1}{\mm}$ and compute the sample mean of the flux objective.
Here, and in the remainder of this manuscript, we use the notation $\langle \cdot \rangle$ to denote that we computed the sample average across coil perturbations for the quantity of interest (such as quadratic flux, violation of quasi-symmetry, or particle confinement).
We recall that the device has a major radius of approximately $R=\SI{1}{\meter}$ and that $\sigma=\SI{1}{\mm}$ implies an expected pointwise error of $\sqrt{\pi}\sigma \approx 0.002 R$. \rev{For devices with different major radius $R$, the coil error magnitude would be scaled linearly to maintain the same relative quadratic flux \ref{eqn:fluxobjective}.}

As expected, for very small perturbations the deterministic minima outperform the stochastic minima. 
If this was not the case, this would indicate that the deterministic algorithm was stuck in a poor local minimum.
We only observe an advantage for stochastic optimization over deterministic optimization once the perturbation magnitude is close to the one used for the stochastic optimization.
However, the difference is not large: for the QA[24] and QA+Well[24] configurations the mean is reduced by $\sim\!13\%$ and $\sim\!10\%$ respectively, and the improvement is smaller still for the shorter coils.

\begin{figure}[H]
    \begin{center}
        \includegraphics[width=\textwidth]{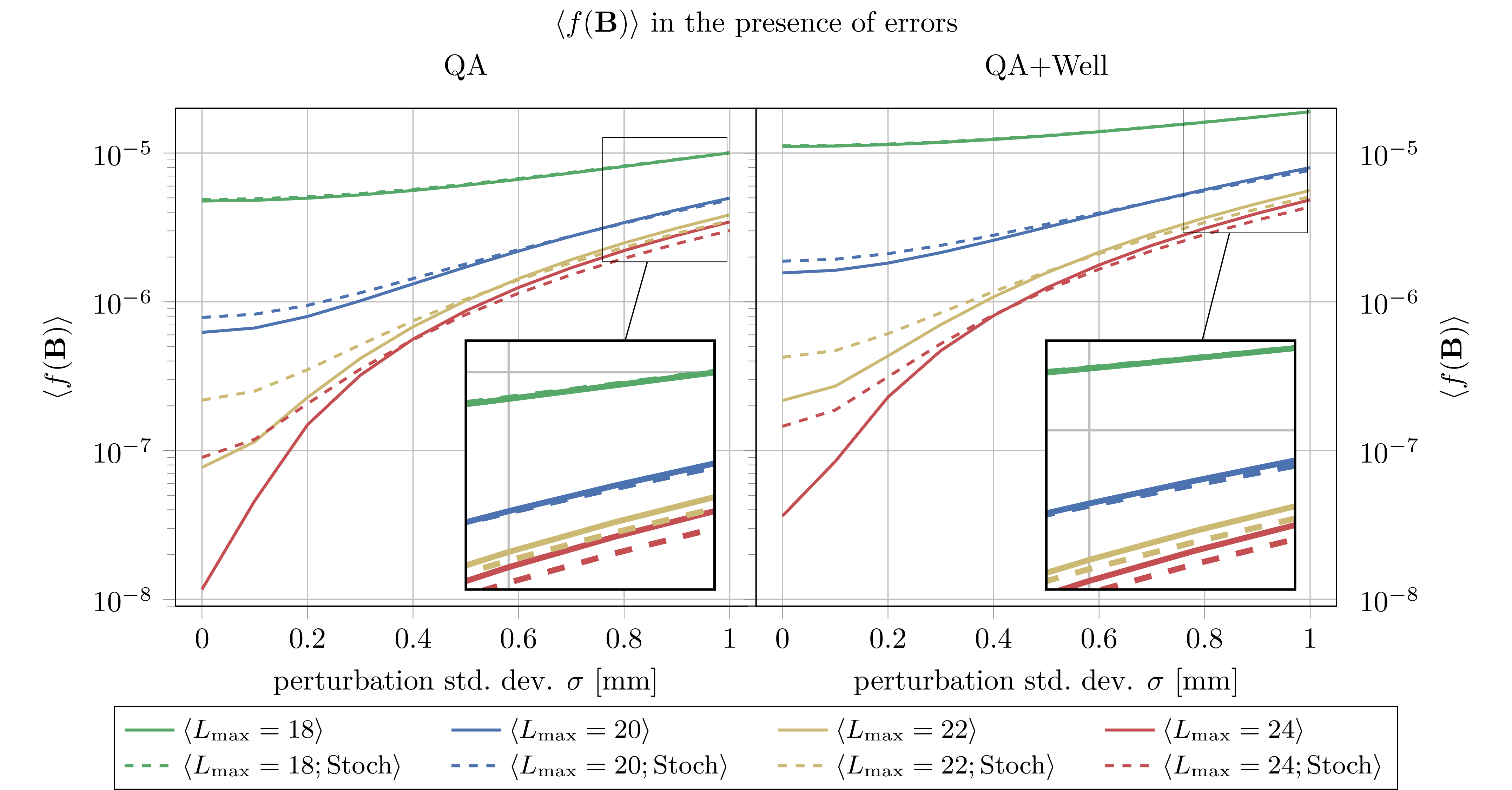}
    \end{center}
    \caption{Impact of coil errors of different magnitude $\sigma$ (shown on the $x$-axis) on the average flux $\langle f(\Bb) \rangle \equiv \Eb_{\zetab}[f(\Bb(\cb, \zetab))]$ for deterministic and stochastic coil designs. The stochastic coil optimization used an error model with $\sigma=\SI{1}{\mm}$. The magnified detail highlights the improved performance of coils from stochastic optimization in the presence of errors, though this improvement is small.}\label{fig:stoch-errorimpact}
\end{figure}

We recall that the motivation behind the flux objective in~\eqref{eqn:fluxobjective} is that if $\Bb\cdot\nb=0$ everywhere on the surface, then the field generated by the coils matches the target field everywhere inside the volume surrounded by the surface.
In particular, it means that the field induced by the coils has the same confinement and quasi-symmetry properties.
However, we are not guaranteed that there is a strict correlation between the flux objective and the physics properties for non-zero values of $f(\Bb)$.
This means that we need to evaluate the quasi-symmetry and the confinement performance of the perturbed coils to study to what extent an improvement of $\langle f(\Bb) \rangle$ translates to improvements in the physical quantities of interest.

\begin{figure}[H]
    \begin{center}
        \includegraphics[width=\textwidth]{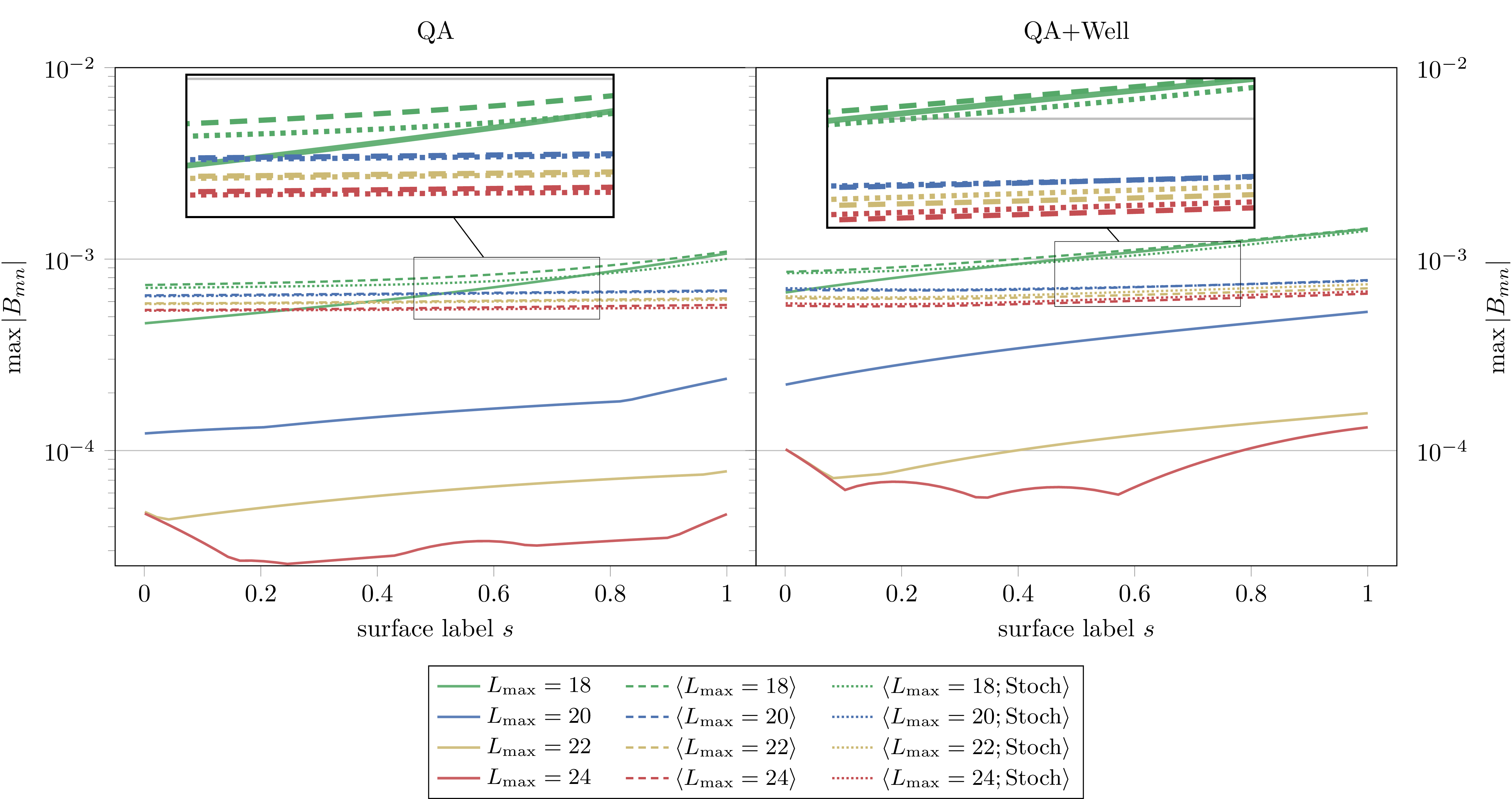}
    \end{center}
    \caption{Sample average of the largest quasi-symmetry breaking modes for $\sigma=\SI{1}{\mm}$. Solid lines: coils from deterministic optimization built exactly. Dashed lines: sample average for coils from deterministic optimization with perturbations. Dotted lines: sample average for coils from stochastic optimization with perturbations.}
    \label{fig:stoch-bmn}
\end{figure}

To study this, we compute a measure of quasi-symmetry and particle confinement for 128 different coil perturbations and compare the performance to that achieved when building and placing the coils obtained from deterministic optimization exactly.
To quantify quasi-symmetry, we recall that a magnetic field is quasi-axisymmetric if the field strength is independent of the angle $\theta$ when expressed in terms of Boozer angles $\phi$ and $\theta$~\cite{Helander2014}.
Thus we can quantify the deviation from quasi-symmetry as follows:
for each perturbed coilset, 
with currents scaled so the average field strength is $\SI{1}{\tesla}$,
we compute surfaces in Boozer angles for the magnetic field using~\cite{giuliani2022}, and perform a Fourier transform of the field strength. That is, we express $|\Bb(s,\theta,\phi)| = \sum_{m,n} B_{m,n}(s) \cos(m \theta - n \phi)$.
For each flux value $s$, we report the largest symmetry-breaking component, i.e. $\max_{n\neq 0, m} |B_{m,n}(s)|$,  and then compute the sample average over all coil perturbations.
The result is shown in Figure~\ref{fig:stoch-bmn}.
The coil errors result in a significant departure from quasi-axisymmetry.
For most, but not all of the configurations, the perturbed stochastic minimizers have slightly better quasi-symmetry than the perturbed deterministic configurations.
We emphasize, however, that this improvement pales in comparison to the significant departure from quasi-symmetry due to coil errors.

\begin{figure}[H]
    \begin{center}
        \includegraphics[width=\textwidth]{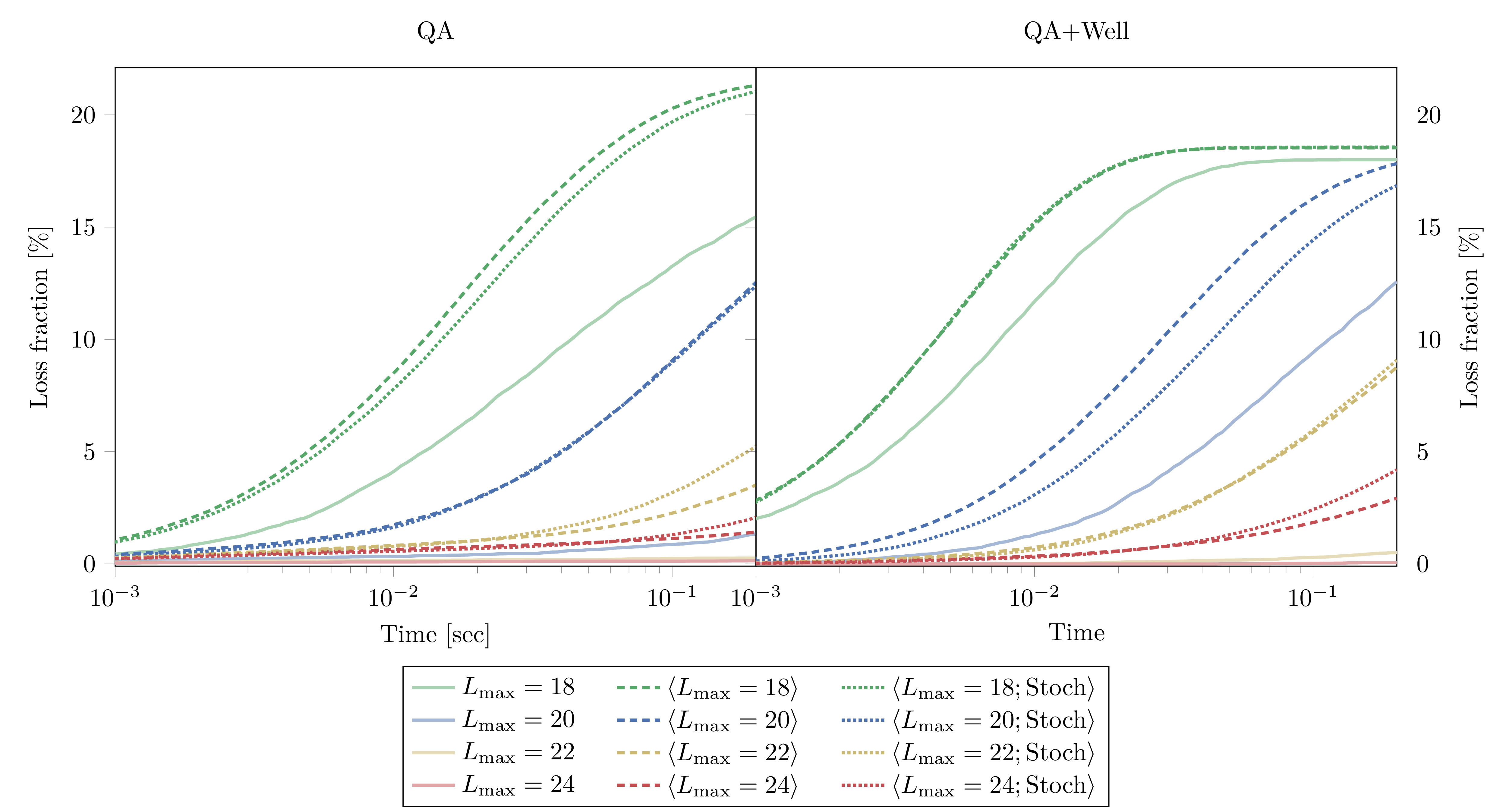}
    \end{center}
    \caption{Confinement of alpha particles spawned on the $s=0.25$ surface. Solid lines: coils from deterministic optimization built exactly. Dashed lines: sample average for coils from deterministic optimization with perturbations. Dotted lines: sample average for coils from stochastic optimization with perturbations.}
    \label{fig:stoch-conf}
\end{figure}

Finally, we evaluate the ability of the fields to confine highly energetic particles. To this end, we scale the device to approximately match the ARIES-CS \cite{najmabadi_aries_cs_2008} reactor:
we choose a minor radius of $\SI{1.7}{\meter}$ and an average field strength on the outer surface of $\SI{5.86}{\tesla}$.
We spawn alpha particles at the surface with half radius and initialize them isotropically in velocity space with 3.5MeV energy, and
numerically compute their collisionless guiding-center trajectories with the SIMSOPT code~\cite{simsopt}.
Again we consider the case in which the coils obtained from deterministic optimization were built and placed perfectly as the reference case, which we compare with a sample average over 128 perturbed coil configurations for both the deterministic and stochastic optimization case.
The losses are shown in Figure~\ref{fig:stoch-conf}.
We make two main observations:
firstly, compared to the case when coils are built exactly (solid lines), the perturbed coils
have significantly higher energetic particle losses. And secondly, the coils found by
stochastic optimization do not systematically outperform those from
deterministic optimization. For the shorter coils, stochastic optimization
seems to result in slightly better confinement, but the opposite is true for
the longer coils. The conclusion here is that when deterministic coil design is
done in a way that avoids being stuck in poor local minima, stochastic
optimization does not yield a substantial benefit for the classical two-stage coil
design problem, at least for the vacuum fields we have considered in this work.

\subsection{Discussion of the impact of stochastic optimization}

\begin{figure}
    \begin{center}
        \resizebox{\textwidth}{!}{%
            \includegraphics[height=4cm]{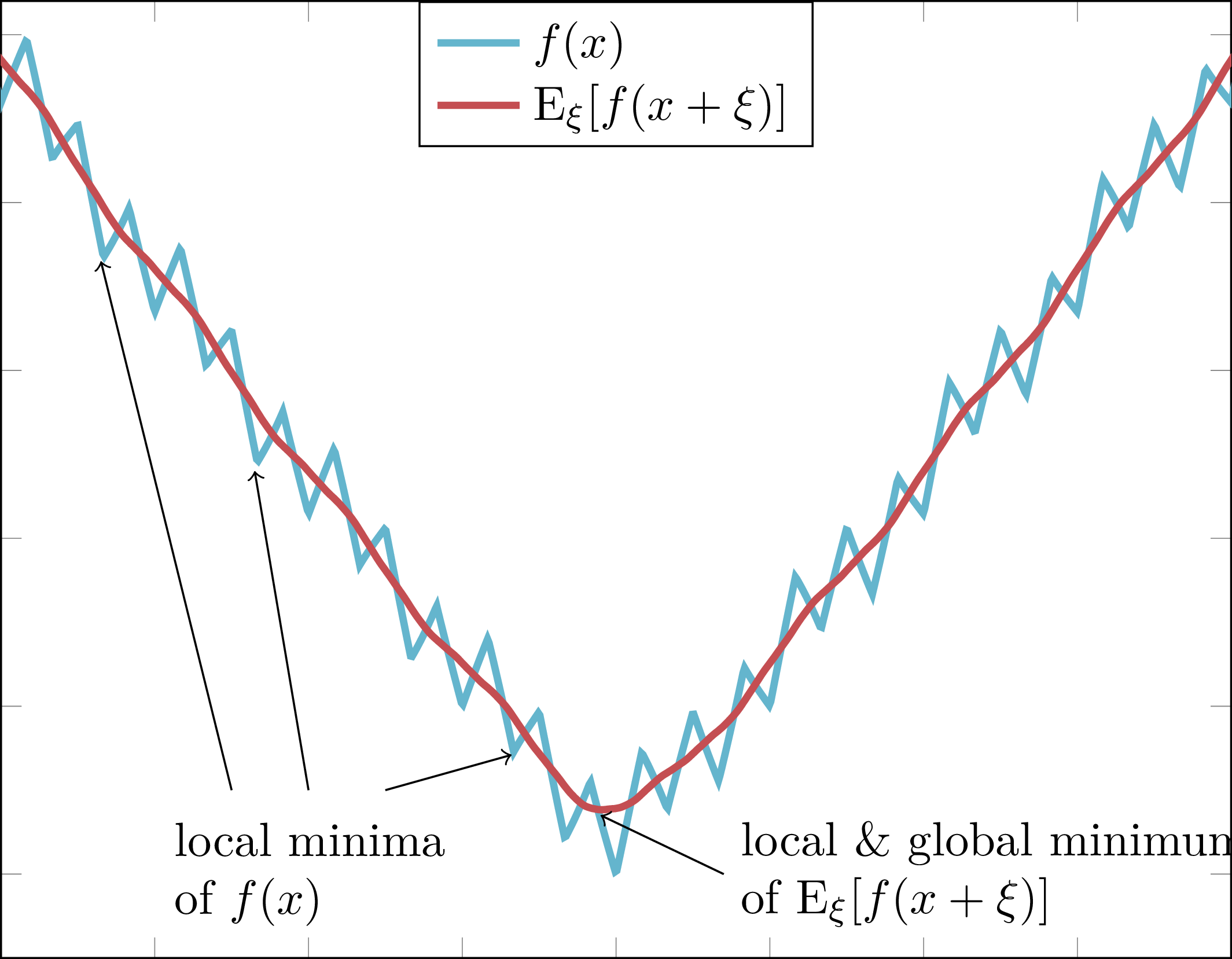}
            \includegraphics[height=4cm]{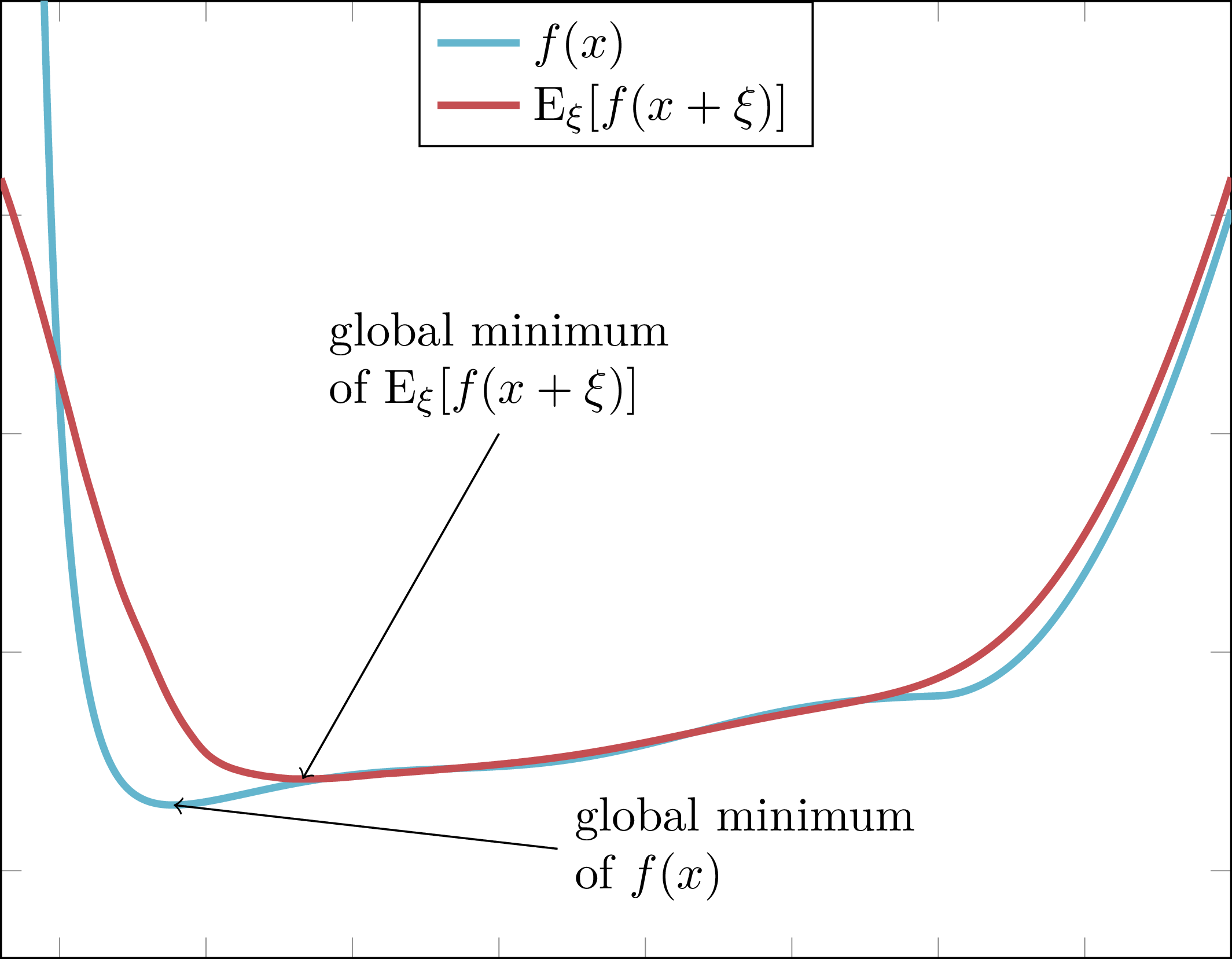}
        }
    \end{center}
    \caption{Illustration of effects of stochastic optimization. Shown \rev{on the left} is the fact that stochastic optimization tends to smooth objective functions and \rev{on the right} that it tends to find minimizers that are more stable, i.e., they lie in wider basins of the original function. }\label{fig:stochastic-optimization-motivation}
\end{figure}

Stochastic optimization has received significant attention over the last few years in the stellarator design community. We focus on the common case of risk-neutral stochastic optimization, i.e., we consider expectations over the random parameter distribution. In order to put our results into context with other work, we start by recalling two main ways in which stochastic optimization impacts objective functions and their minimizers. 
As an illustrative example, we assume that the stochastic objective is of the form $\Eb[f(\xb + \xib)]$ for a Gaussian random vector $\xib$.
In this case, stochastic optimization can be interpreted as a convolution with a Gaussian kernel. In particular, stochastic objectives typically are smooth versions of their deterministic counterparts.  This has two main implications, which are demonstrated in Figure~\ref{fig:stochastic-optimization-motivation}.
Firstly, this smoothing may
reduce the number of local minima and maxima.
Secondly, deterministic minima that are very close to much larger function values may not be minima of the stochastic objective as the stochastic objective takes into account a neighborhood around each point. This is why minimizers computed from stochastic formulations tend to perform robustly with respect to random perturbations. 
%
Since we consider several different initial guesses for the optimization and pick the best performing local minimum, any difference between deterministic and stochastic optimization in our results are likely due to the second effect above.

Stochastic optimization was previously applied to the problem of coil
design in~\cite{lobsien_stellarator_2018}, where a stochastic version of
the ONSET code is used to find coils for the W7-X experiment and it is observed
that the objective value is reduced by up to 21\% when comparing stochastic
to deterministic minimization. The main reason for this significant improvement
appears to be due to the deterministic minimizer getting stuck in a poor local
minimum.
We can see that this is the case, since even in the absence of errors, the minima obtained from stochastic optimization outperform those obtained from deterministic optimization by up to 19\%---implying that the latter can not be the global minimum (see Figure~1 in
\cite{lobsien_stellarator_2018} and compare the values given for $f(x_0)$).
Hence, it seems to be the first of the two effects in
Figure~\ref{fig:stochastic-optimization-motivation} that has a significant impact.

We postulate that the reason why ONSET struggles with local minima is a
combination of the choice of objective and optimization algorithm: ONSET
contains several non-smooth contributions to the objective (such as the maximal field
error), and uses a derivative-free optimization algorithm. In contrast, the
FOCUS-like formulation that we use here only involves smooth functions and we use a derivative-based algorithm, which appears to lead to fewer problems with local minima.
In fact, we note that the different local minima that we find as a result of using different initialization all perform rather similarly well and, while different, are visually similar too.  

More recently, in~\cite{wechsung2021} stochastic optimization was applied to
single-stage coil design, that directly optimized coils for quasi-symmetry near
the magnetic axis, as presented in~\cite{giuliani2020}. When performing
deterministic optimization for the single-stage quasi-symmetry objective in
\cite{giuliani2020}, we are able to achieve objective values of $<10^{-10}$.
However, even small perturbations result in objective values that
are several orders of magnitude larger. This suggests that we are in the second
situation of Figure~\ref{fig:stochastic-optimization-motivation}, and that the
improvement is not due to the avoidance of local minima, but due to stochastic minimization avoiding unstable locations in parameter space. 

\section{A posteriori coil error correction}
Much of today's understanding of coil manufacturing errors and their impact can be credited to the Wendelstein 7-X experiment. During the construction of W7-X,
manufacturing errors of the coils were measured after their production and the
position of the coils was adjusted one by
one~\cite{andreeva_influence_2009,bosch_technical_2013} to minimize error-fields to first order.
Motivated by this idea, we now systematically study the ability of a-posteriori
coil adjustments to reduce the impact of coil errors. We will use the error
model from the previous section to simulate coil errors of various
magnitudes and then evaluate how well these errors can be mitigated.

While, once built, the shape of the coil cannot be changed, the coil can still be
shifted, rotated, and its current can be adjusted. There are therefore seven
degrees of freedom per coil that could be optimized: three degrees of freedom
corresponding to translations in each direction, three degrees of freedom
corresponding to rotations around each axis, and one degree of freedom for the
current in the coil. Here, we consider each of the 16 coils independently, as
manufacturing errors might not satisfy stellarator symmetry or field period
symmetry and hence each coil might require a distinct correction.

Denoting, for the $i$th coil, the translation by $\ssb^{(i)}\in \mathbb R^3$ and the rotation angles by $\alphab^{(i)}=(\alpha^{(i)}_1, \alpha^{(i)}_2, \alpha^{(i)}_3)\in \mathbb R^3$, we consider the rigid transformation
\begin{equation}
  \Ob_{\ssb^{(i)}, \alphab^{(i)}}(\xb) = \Rb_\text{yaw}(\alpha^{(i)}_1) \Rb_\text{pitch}(\alpha^{(i)}_2) \Rb_\text{roll}(\alpha^{(i)}_3) \xb + \ssb^{(i)}
\end{equation}
with the rotation matrices
\begin{equation}
    \begin{aligned}
        \Rb_\text{yaw}(\alpha) &=  \begin{bmatrix}
            \cos(\alpha) & -\sin(\alpha) & 0 \\
            \sin(\alpha) & \cos(\alpha) & 0 \\
            0 & 0 & 1
    \end{bmatrix},\\
    \Rb_\text{pitch}(\alpha) &=\begin{bmatrix}
            \cos(\alpha) &  0 & \sin(\alpha) \\
            0 & 1 & 0 \\
            -\sin(\alpha) & 0  & \cos(\alpha)
    \end{bmatrix},\\
        \Rb_\text{roll}(\alpha)&=\begin{bmatrix}
            1 & 0 & 0 \\
            0 & \cos(\alpha)& -\sin(\alpha) \\
            0 & \sin(\alpha) & \cos(\alpha)
    \end{bmatrix}.
    \end{aligned}
\end{equation}

We denote the magnetic field induced by the corrected coils by
\begin{equation}
    \Bb(\cb, \zetab, \{(\ssb^{(i)}, \alpha^{(i)}, \delta I^{(i)})\}_{i=1}^{16}) = \Bb(\{\Ob_{\ssb^{(i)}, \alphab^{(i)}} (\Gammab^{(i)}(\cb) + \Xib^{(i)}(\zetab))\}_{i=1}^{16}, \{I^{(i)}(\cb) + \delta I^{(i)}\}_{i=1}^{16})
\end{equation}
and then for a given design $\hat \cb$ (obtained from deterministic optimization in this case) and measured manufacturing errors $\hat \zetab$, we solve
\begin{equation}\label{eqn:apostioptimproblem}
    \begin{aligned}
        \underset{\{(\ssb^{(i)}, \alpha^{(i)}, \delta I^{(i)})\}_{i=1}^{16}}{\mathrm{minimize}}\quad &   f\big(\Bb(\hat\cb, \hat\zetab, \{(\ssb^{(i)}, \alpha^{(i)}, \delta I^{(i)})\}_{i=1}^{16})\big),\\
        \text{subject to}     \quad& \rev{\begin{cases}\text{the curves } \{\Ob_{\ssb^{(i)}, \alphab^{(i)}} (\Gammab^{(i)}(\cb) + \Xib^{(i)}(\zetab))\}_{i=1}^{16}\\
                                    \text{satisfying the distance constraints.}
                                   \end{cases}}
    \end{aligned}
\end{equation}
Since coil length and curvature are not affected by rigid transformations, all other geometric constraints considered in the optimization \eqref{eqn:optimproblem} or its deterministic counterpart are satisfied automatically.
We note that our approach differs from the one employed for W7-X in a few
aspects. Firstly, all coils are adjusted at once, rather than sequentially.
Secondly, the objective that we minimize is the same as the one used during
coil design, rather than a separate error field objective, and lastly, in
addition to considering coil shifts and rotations, we also consider current
adjustments as a degree of freedom that can be used to adjust the magnetic
field. To better understand the impact of the current in addition to the placement correction,
we consider three variants of this a-posteriori optimization, which we refer to as the correction levels (CL):
\begin{enumerate}[label=(CL\arabic*), align=left, leftmargin=*]
    \item Only optimize for the position and orientation of the coils. The resulting optimization problem has $16\cdot 6=96$ degrees of freedom.
    \item CL1 plus a current correction per modular coil group. This is motivated by the fact that usually each set of four nominally identical modular coils is connected in series to the same power supply, so adjusting the current on a per-coil basis is more complicated. The resulting optimization problem has $16\cdot 6 + 4=100$ degrees of freedom.
    \item CL1 plus a current correction for each individual coil. The resulting optimization problem has $16\cdot 7=112$ degrees of freedom. This corresponds to the situation in~\eqref{eqn:apostioptimproblem}.
\end{enumerate}

\begin{figure}[bth]
    \begin{center}
        \includegraphics[width=0.32\textwidth]{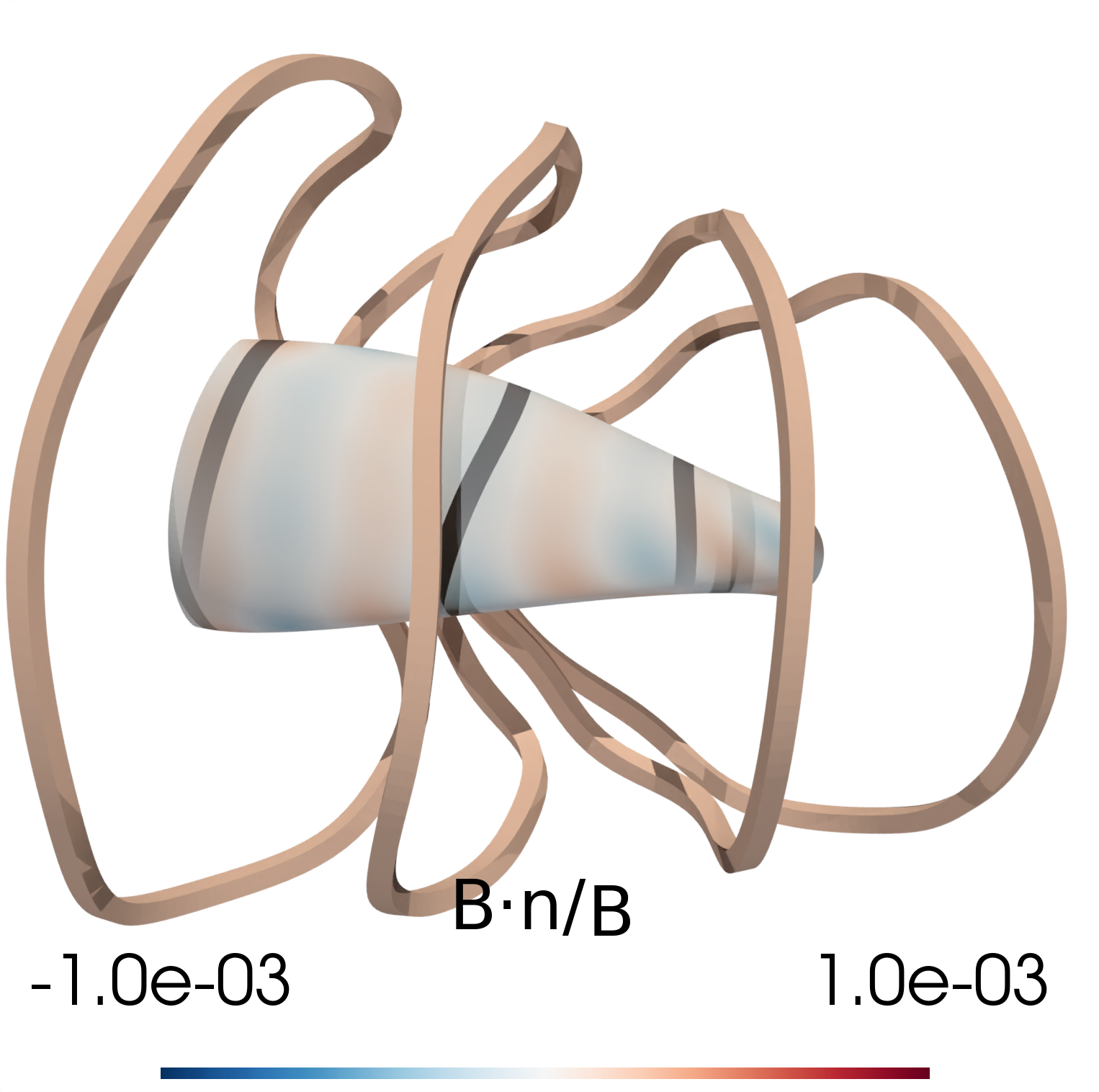}\hfill
        \includegraphics[width=0.32\textwidth]{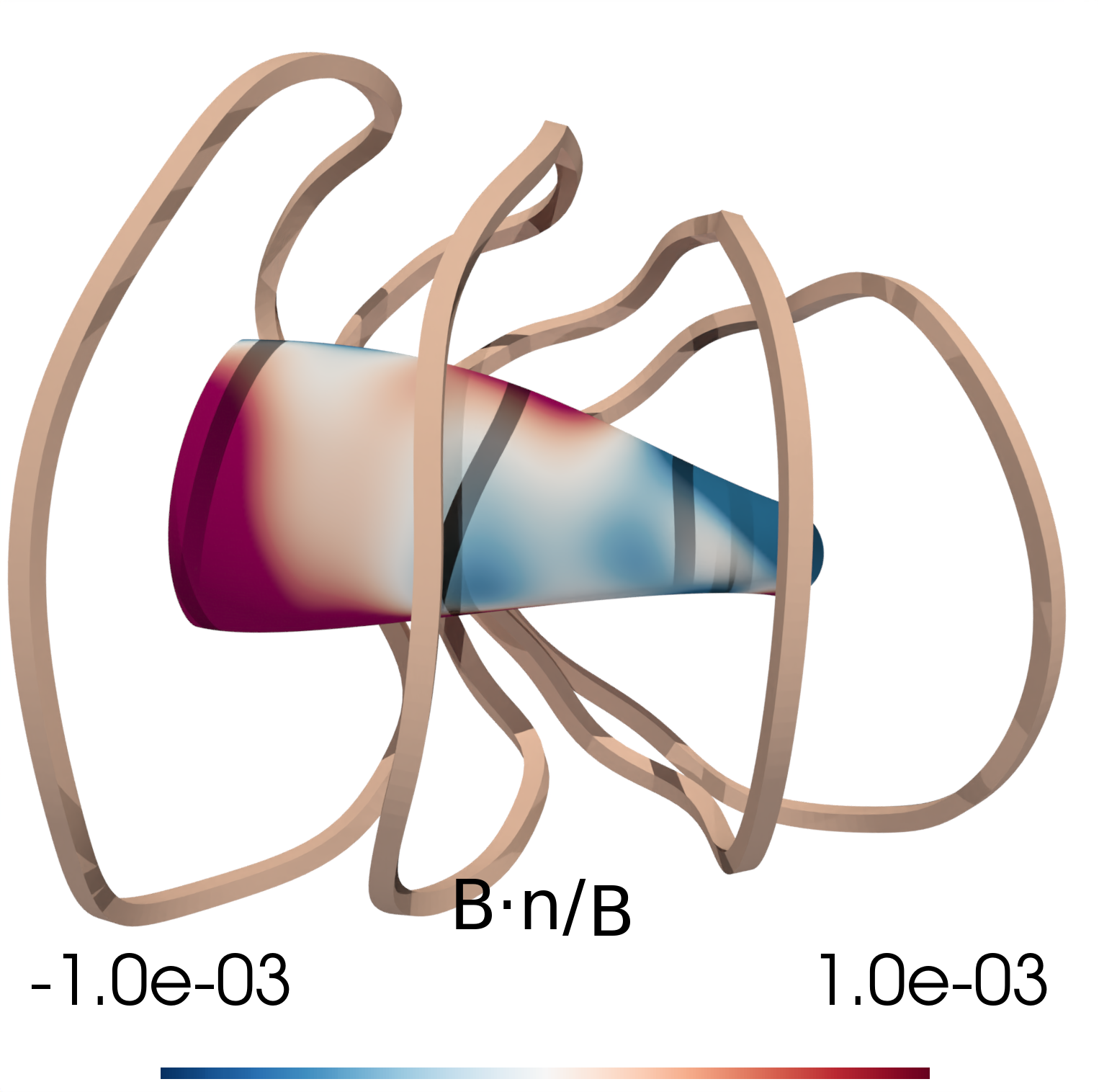}\hfill
        \includegraphics[width=0.32\textwidth]{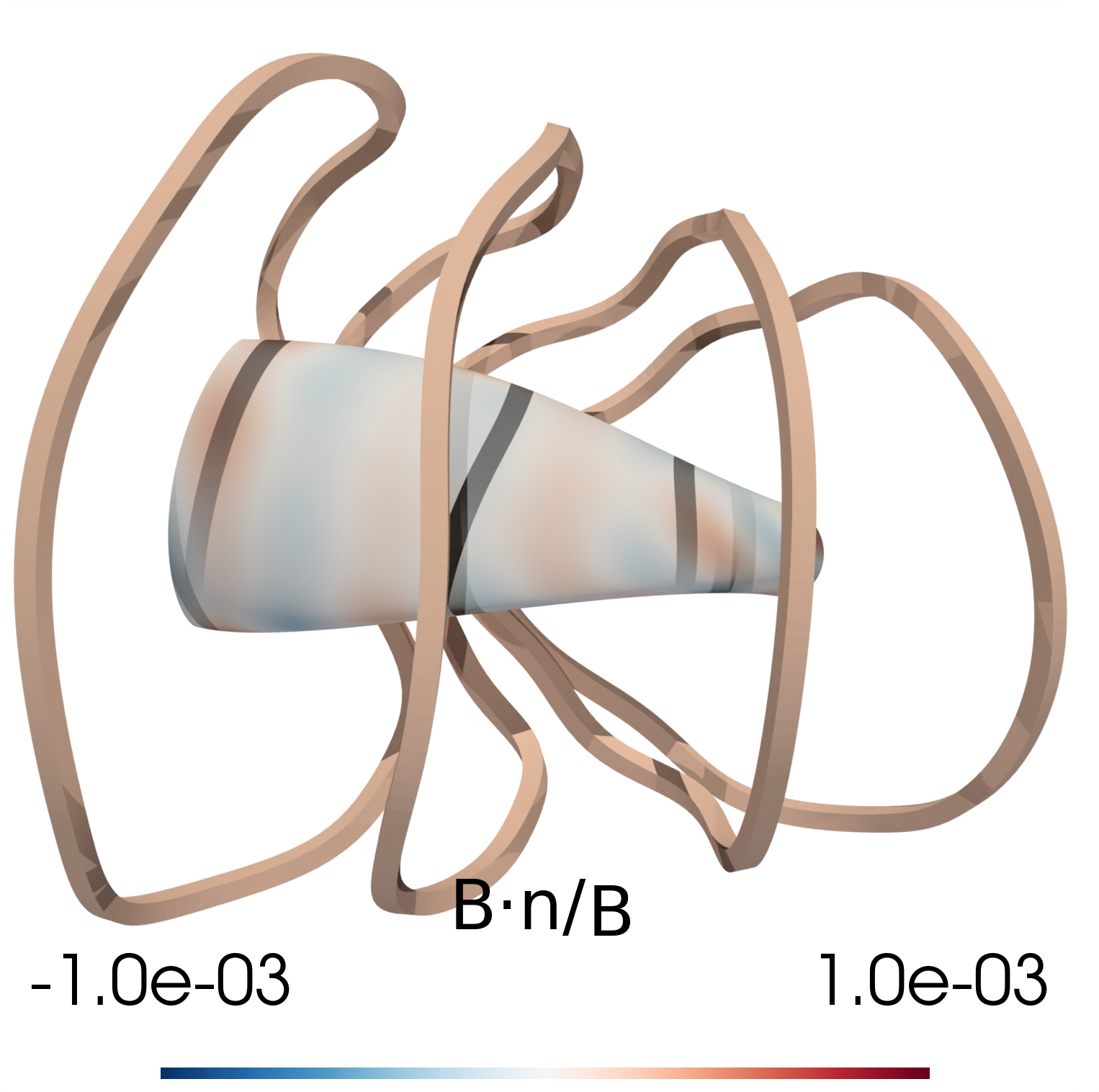}
    \end{center}
    \caption{Left: QA[22] coils. Middle: QA[22] coils with random draw from the error model ($\sigma=\SI{2}{\mm}$). Right: QA[22] coils with error and CL3 correction. Note that the differences between the coils are too small to be visible. Coils are modelled as line filaments and the finite cross section is for visual purposes only. The color on the surface indicates ${\Bb\cdot\nb}/{|\Bb|}$. We observe that the a-posteriori corrections are able to almost entirely undo the impact of the coil errors.  }\label{fig:coils-errorcorrection}
\end{figure}

In Figure~\ref{fig:coils-errorcorrection}, we show an example of the impact of coil errors and how well they can be corrected. While in the figure we only show four coils, all 16 coils may have different errors, and each coil position is corrected by solving the optimization problem \eqref{eqn:apostioptimproblem} over rigid body motions and currents for each coil. As can be seen, the normal flux increases significantly due to errors, but that increase can be almost entirely eliminated by the coil position, orientation, and current adjustments. 

\begin{figure}[H]
    \begin{center}
        \includegraphics[width=\textwidth]{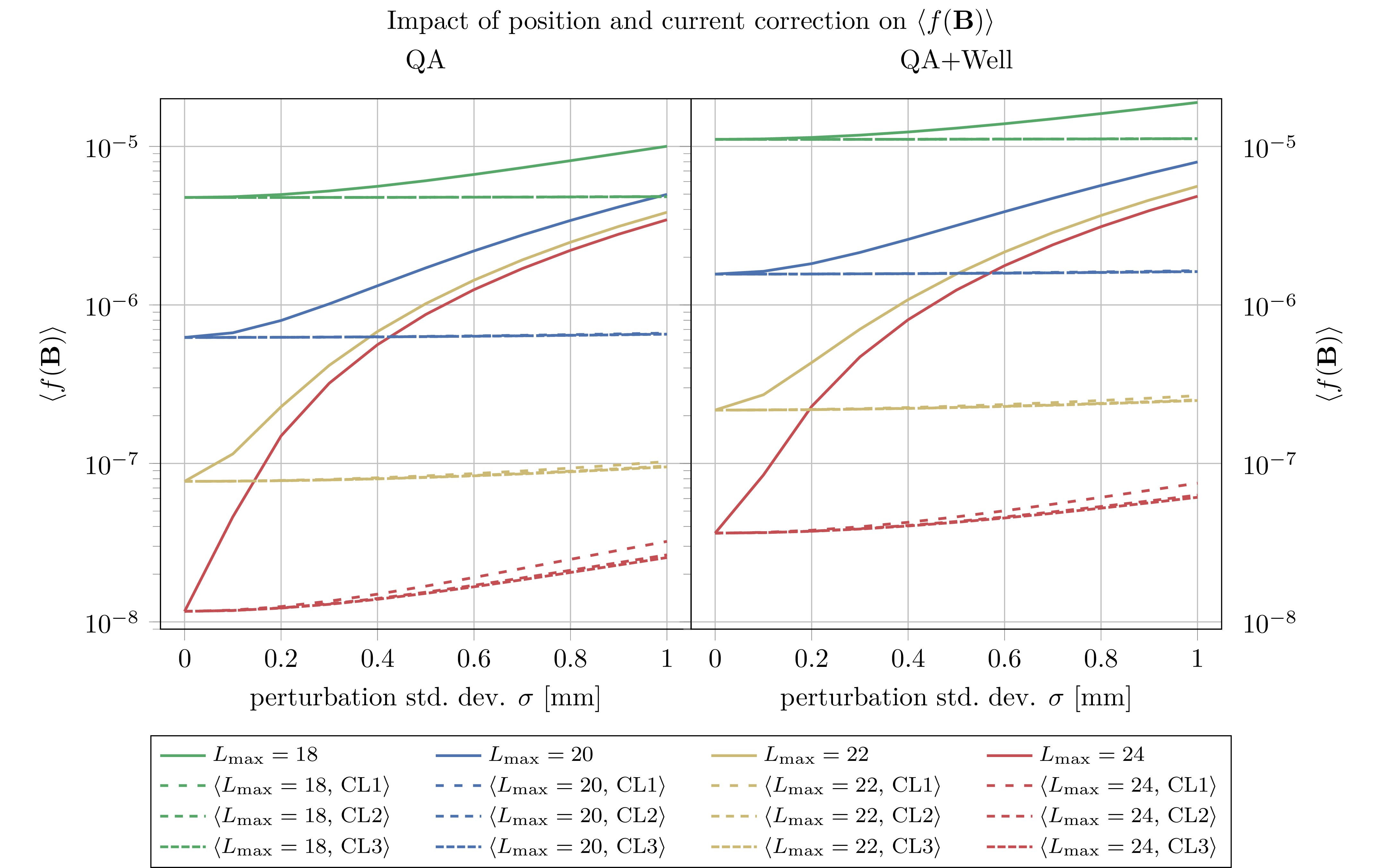}
    \end{center}
    \caption{Impact of coil errors with different $\sigma$ ($x$-axis) on $\langle f(\Bb)\rangle $ for the deterministic minimizers both without correction (solid) and with the three different a posterior correction levels (dashed). It can be seen that the corrections substantially mitigate the degradation of the flux objective.}\label{fig:flux-errorcorrection}
\end{figure}

\begin{remark}
\rev{Note that when building a stellarator experiment, one may not be able to wait until all the coils have been manufactured to start placing them. To address such situations, one can formulate a sequence of similar optimization problems that adjust coils or groups of coils as they arrive from manufacturing. The all-at-once optimization problem considered here can be viewed as a best case scenario in which measurements of all coils are available.}
\end{remark}
To systematically evaluate to what degree accurately measured manufacturing errors can be corrected, we again draw 128 random perturbations to be added to the deterministic minimizer.
Then, for each set of perturbations, we solve the optimization problem \eqref{eqn:apostioptimproblem} for each of the correction levels in order to minimize the flux objective.
We repeat this procedure for multiple perturbations and error magnitudes and plot the resulting sample averages in Figure~\ref{fig:flux-errorcorrection}.
\rev{We find that} coil manufacturing errors can \rev{consistently} be corrected to a large degree by adjusting coil placement and currents.
Even when keeping the currents fixed and only adjusting for the position of the coils, the objective value only increases slightly, i.e., most of the manufacturing error can be corrected by a slight correction of the coil position. \rev{In particular for the shorter coils, (CL1), (CL2), and (CL3) yield near identical results. For the longest coils and largest errors considered here, adding group-wise current corrections (CL2) improves the flux objective by $\sim18\%$, and adding individual current corrections (CL3) reduces the error by another $\sim4\%$.} 

\begin{figure}[H]
    \begin{center}
        \includegraphics[width=\textwidth]{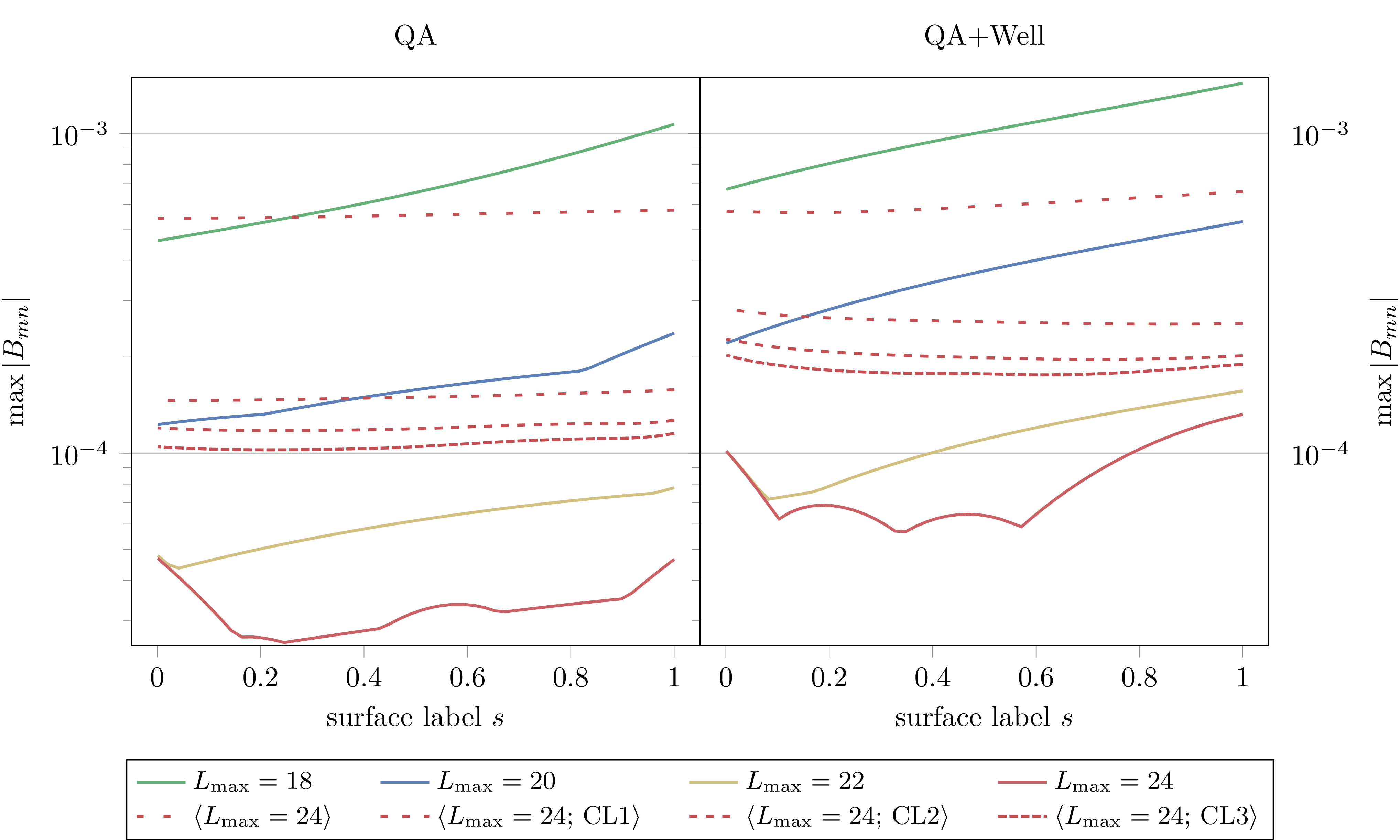}
    \end{center}
    \caption{Sample average for errors with $\sigma=\SI{1}{\mm}$ of the largest quasi-symmetry breaking modes as a function of surface label $s$ ($x$-axis). Shown are the results for the longest coilset without correction and with all three correction levels (dashed lines). As reference, we also show the results without errors (solid lines) for different coil lengths $\lmax$.}
    \label{fig:corr-bmn}
\end{figure}

To study how well improvement in the flux objective translates to better physics performance, we compute the quasi-symmetry violation and the energetic particle confinement properties for 128 perturbed stellarators and for each correction level.
For conciseness, we focus on the case $\lmax=24$ and $\sigma=\SI{1}{\mm}$ for these computations.
In Figure~\ref{fig:corr-bmn} we can see that the position correction (CL1) yields a significant improvement of quasi-symmetry. 
Adding the current correction in CL2 and CL3 then yields incremental additional improvement.

\begin{figure}[H]
    \begin{center}
        \includegraphics[width=\textwidth]{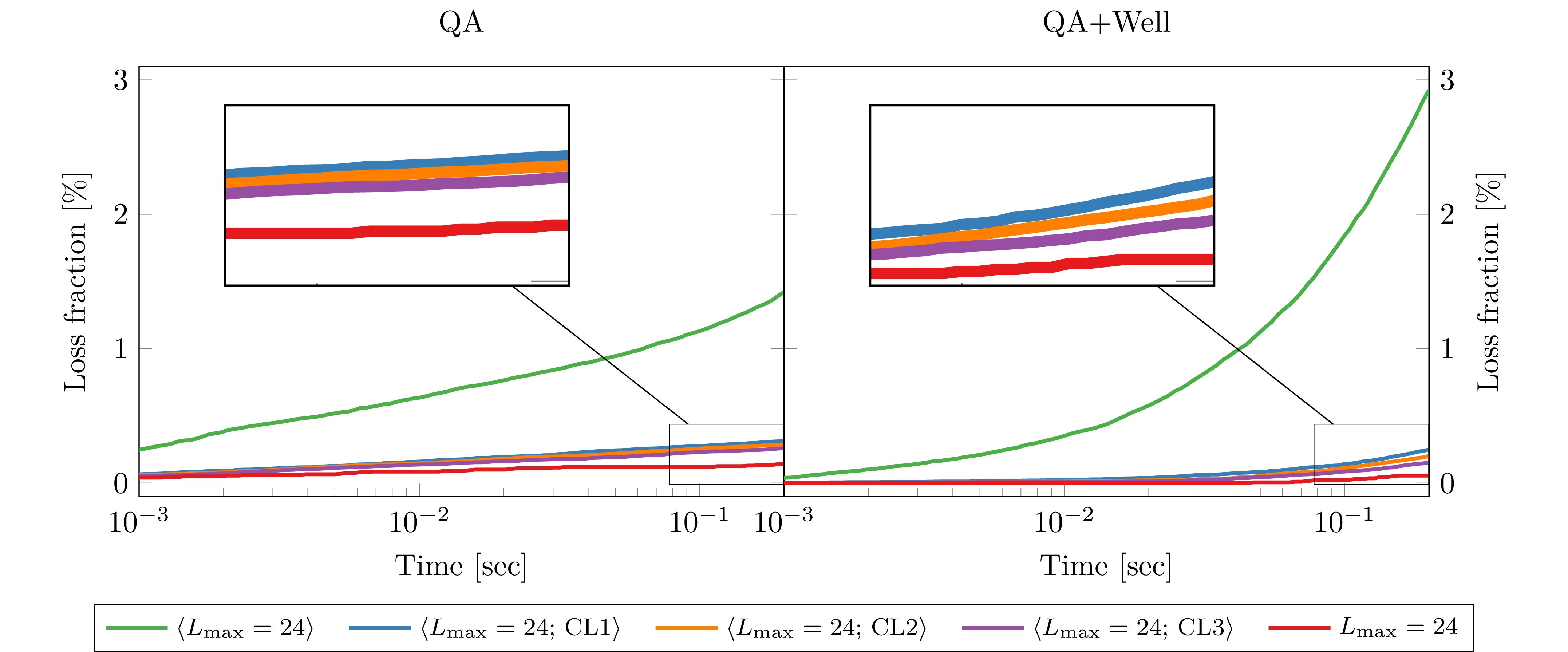}
    \end{center}
    \caption{Particle confinement of alpha particles spawned on the $s=0.25$ surface for the exactly built coils (red), for 128 perturbed coils without correction (green), and with all three correction levels (blue, orange, purple). 
        As the correction level increases, particle losses decrease.
    }
    \label{fig:corr-conf}
\end{figure}

Finally, \rev{using the same setup as described at the end of Section \ref{sec:stoch-num}}, we consider alpha particle confinement in Figure~\ref{fig:corr-conf}.
\rev{Without corrections, we lose $\sim 1.42\%$ and $\sim 2.92\%$ of all particles respectively after $\SI{200}{ms}$.}
We observe that particle losses are reduced significantly when adjusting the position of the coils, \rev{almost to the magnitude obtained by coils without manufacturing errors} .  Furthermore, \rev{minor} improvement can be obtained by also adjusting their currents.
For the highest correction level, we lose only $\sim\! 0.26\%$ for the QA, and $\sim\! 0.17\%$ of the QA+Well configuration, a fairly modest increase compared to $\sim\!0.14\%$ and $\sim\!0.06\%$ respectively for the perfectly built and placed coils. 

\section{Discussion and Conclusion}
In this work we have investigated the impact of coil manufacturing errors on stellarator coils generating magnetic fields that are quasi-symmetric to a very high accuracy.
We have shown that fairly small errors have considerable impact on the confinement properties of the induced magnetic fields, and that stochastic optimization does not enable substantial mitigation of coil errors.
On the other hand, if we assume that coil errors can be measured accurately, we showed that particle confinement can be improved significantly by solving a separate optimization problem for an optimal adjustment of the coils.
This adjustment includes shifts and rotations of the coils, and optionally can include changes of the coil currents.
While we used our error model to assess this approach, we highlight that no such model is required when applying this approach during construction:
the initial coil design can be performed without consideration of errors, and the a-posteriori optimization corrects for specific manufacturing errors after they were measured.

While, in practice, errors cannot be perfectly measured and coils cannot be placed perfectly, \rev{our study highlights the efficacy of a-posteriori coil adjustments as a tool for dealing with coil manufacturing errors.}

\section*{Code availability}
The code used for the simulations and analysis in this manuscript is publicly available at 
\begin{center}
    \small
    \url{https://github.com/florianwechsung/Stage-II-Optimization-With-Perturbations}
\end{center}
and has been archived at~\cite{zenodo_stage_two}.

\section*{Acknowledgments}
The authors would like to thank the SIMSOPT development team and Thomas Sunn Pedersen for his insight on the construction and coil errors of W7X.
In addition, this work was supported in part through the NYU IT High Performance Computing resources, services, and staff expertise.
This work was supported by a grant from the Simons Foundation (560651). AG is partially supported by an NSERC (Natural Sciences and Engineering Research Council of Canada) postdoctoral fellowship.  In addition, AC and FW are supported by the United States National Science Foundation under grant No.\ PHY-1820852, and AC is supported by the United States Department of Energy, Office of Fusion Energy Sciences, under grant No. DE-FG02-86ER53223.

\printbibliography
\end{document}